\documentclass[conference]{IEEEtran}
\IEEEoverridecommandlockouts

\usepackage{cite}
\usepackage{amsmath,amssymb,amsfonts}
\usepackage{algorithmic}
\usepackage{graphicx}
\usepackage{textcomp}
\usepackage{xcolor}
\usepackage{booktabs}
\usepackage{multirow}
\usepackage{graphicx}
\usepackage{xcolor}

\usepackage{tikz}

\usetikzlibrary{positioning, arrows.meta, calc, shapes.geometric}

\def\BibTeX{{\rm B\kern-.05em{\sc i\kern-.025em b}\kern-.08em
    T\kern-.1667em\lower.7ex\hbox{E}\kern-.125emX}}
\begin{document}

\title{CTC-Seeded Token Edit Refinement for Non-Autoregressive Speech Recognition}

\author{
\IEEEauthorblockN{Wanting Huang}
\IEEEauthorblockA{
\textit{Department of Computer Science} \\
\textit{University of Iowa} \\
Iowa City, IA, USA \\
wanting-huang@uiowa.edu
}
\and
\IEEEauthorblockN{Weiran Wang}
\IEEEauthorblockA{
\textit{Department of Computer Science} \\
\textit{University of Iowa} \\
Iowa City, IA, USA \\
weiran-wang@uiowa.edu
}
}

\maketitle

\begin{abstract}

Non-autoregressive automatic speech recognition (ASR) enables parallel decoding, but many refinement-based methods begin from random, fully masked, or fixed-length token sequences, requiring multiple iterations to reconstruct the complete transcript. We instead formulate ASR decoding as a variable-length edit refinement of a greedy connectionist temporal classification (CTC) hypothesis. An acoustic-conditioned Edit Flow decoder operates directly on the collapsed CTC hypothesis, predicting insertion, deletion, and substitution operations in parallel. The Edit Flow decoder is jointly trained with a CTC model using a continuous-time discrete diffusion loss. During inference, we find that just two edit steps yield substantial Word Error Rate (WER) reductions, and classifier-free guidance (CFG) further enhances recognition quality by focusing the model on audio features. We also constrain edit proposals using CTC confidence to improve accuracy. Finally, ablation studies validate our design choices, while decoder pretraining and pretrained encoder integration yield significant additional performance gains.
\end{abstract}


\section{Introduction}
\label{sec:introduction}

End-to-end automatic speech recognition (ASR) systems are typically based on connectionist temporal classification (CTC~\cite{graves2006ctc}), recurrent neural network transducers (RNN-T~\cite{graves2012rnnt}), or attention-based encoder--decoders (AED~\cite{chan2016las}). While RNN-T and AED models successfully capture dependencies among output tokens using a decoder that acts as an implicit language model, their autoregressive generation inherently limits inference speed. In contrast, CTC imposes a strict conditional independence assumption on the label sequence. Although this degrades accuracy, it permits highly efficient parallel greedy decoding, making CTC an ideal foundation for non-autoregressive (NAR) ASR. To mitigate the accuracy tradeoff, prior NAR 
methods rely on iterative masked prediction, latent-position imputation, or iterative alignment refinement~\cite{higuchi2020maskctc,chan2020imputer,chi2021align} to gradually reconstruct label dependencies and enhance overall recognition quality.


Diffusion models provide a powerful alternative to classical NAR methods by using a principled objective that couples all generation steps. They update all token positions in parallel and offer an accuracy–efficiency trade-off controlled by the number of decoding steps. In typical sequence generation, these iterative decoders start from uninformative states—such as fully random tokens, fully masked sequences, or fixed-length latents~\cite{austin2021structured,ghazvininejad2019maskpredict,li2022diffusion}. However, this from-scratch approach is inefficient for ASR, where greedy CTC already provides an inexpensive and informative first-pass hypothesis. Starting from an uninformative state forces the model to waste evaluations reconstructing tokens CTC has already predicted correctly.

To address this, we adopt a correction-oriented view of NAR decoding. Rather than generating a transcript from scratch, we formulate ASR as a CTC hypothesis-to-transcript edit refinement. The collapsed greedy CTC output supplies a strong initial token sequence. Because this collapsed sequence is completely free of blanks, it is substantially shorter than a frame-level hypothesis, making our method significantly more efficient than alignment-level refinement~\cite{chan2020imputer,chi2021align,wang2021deliberation}. Starting from this shorter sequence allows the decoder to focus solely on correcting residual errors—which we naturally represent as insertion, deletion, and substitution operations. Furthermore, when the initial CTC hypothesis is already close to the reference in edit distance, a small number of parallel correction rounds can effectively replace full-sequence reconstruction.

We implement this approach using an acoustic-conditioned Edit Flow decoder~\cite{havasi2026edit}. By operating directly on the collapsed CTC token sequence, our variable-length edit formulation elegantly avoids padded sequences and separate length predictors. To ensure the decoder makes targeted corrections rather than freely rewriting the hypothesis, we introduce inference strategies designed to strongly condition the refinement process on the underlying acoustic evidence.

In summary, our main contributions are:
\begin{itemize}
\item 
We formulate NAR ASR decoding as a token-level, variable-length refinement of a greedy CTC hypothesis. Accordingly, we train an acoustic-conditioned Edit Flow decoder via a discrete diffusion objective to recover ground-truth transcripts from noisy CTC baselines.

\item 
We introduce a novel inference strategy that gates model-proposed edits based on the CTC model's acoustic confidence. When combined with audio classifier-free guidance (CFG), this targeted approach yields significant accuracy gains and achieves optimal performance in just two parallel editing iterations.

\item 
We design an Edit Flow-specific text pretraining scheme that utilizes deletion, substitution, and insertion corruptions. Furthermore, we conduct extensive ablation studies on the LibriSpeech dataset to analyze the impact of model scale, pretraining effects, and decoding strategies. Our 
system delivers substantial improvements over the first-pass CTC baseline. Specifically, on test-clean/test-other, it reduces the WER from 3.5/7.9 to 2.6/5.8 with the ESPnet encoder (relative reductions of 25.7\% and 26.6\%), and from 2.6/6.1 to 2.0/4.7 with the Whisper medium encoder (relative reductions of 23.1\% and 23.0\%).
\end{itemize}
\section{Related work}
\label{sec:related}

\subsection{Token-level NAR}

Non-autoregressive (NAR) ASR replaces token-by-token generation with parallel prediction. While CTC directly realizes this paradigm, its framewise conditional independence assumption inherently limits its ability to model dependencies among output tokens. To address this, several methods iteratively refine a CTC output. For example, Mask-CTC~\cite{higuchi2020maskctc} starts with a greedy CTC hypothesis, masks low-confidence tokens, and reconstructs those positions using bidirectional context. Its extension, Improved Mask-CTC~\cite{higuchi2021improvedmaskctc}, introduces partial-target length prediction, allowing the model to explicitly handle insertion and deletion errors during refinement. Beyond masking, other approaches explore alternative formulations: Insertion-based ASR~\cite{fujita2020insertionasr} constructs transcripts iteratively through insertion operations, while CASS-NAT~\cite{fan2023cassnat} extracts token-level acoustic embeddings from CTC alignment spans to predict the final output sequence in parallel. \cite{futami2022pcmlm} incorporates phonetic information in token-level masked prediction.

\subsection{Alignment-level refinement}


Another class of NAR methods performs iterative refinement based on the CTC alignment, which includes blank tokens to indicate non-emission. For instance, Imputer~\cite{chan2020imputer} iteratively fills 
alignment positions, using dynamic programming to approximately marginalize over monotonic alignments and generation orders. Similarly, Align-Refine~\cite{chi2021align} repeatedly updates a frame-level CTC alignment before collapsing it into a final transcript. In this framework, changes in output length naturally arise through updates to blanks and repeated labels, but the evolving generation state remains a complete, frame-level alignment. Building on these concepts, non-autoregressive deliberation~\cite{wang2021deliberation} applies a related alignment-refinement strategy to first-pass streaming RNN-T outputs, while Streaming Align-Refine~\cite{wang2022streamingalignrefine} restricts the available context to support low-latency recognition.


\subsection{Diffusion-based ASR}


Diffusion-based ASR models update all token positions in parallel within each denoising step, naturally exposing an accuracy--efficiency trade-off determined by the number of sequential refinement steps. Early approaches like TransFusion~\cite{baas2022transfusion} progressively transform a random character sequence into a transcript conditioned on pretrained speech representations. Similarly, cross-modality diffusion conditions a discrete denoising process on acoustic features, leveraging fast sampling to reduce inference costs~\cite{yeh2024crossmodal}. dLLM-ASR~\cite{tian2026dllmasr} utilizes masked diffusion~\cite{nie2025llada} with a confidence-based, variable number of denoising steps and an optional CTC prior initialization. MDM-ASR~\cite{yen2026mdmasr} performs masked diffusion~\cite{puvvada2024less} initialized from a strong pretrained encoder--decoder model, incorporating iterative self-correction training and confidence-based sampling. Audio-conditioned diffusion LLMs~\cite{wang2025audio} further explore diffusion-based decoding for ASR and deliberation, employing random masking, low-confidence masking, and semi-autoregressive strategies. Drax defines an audio-conditioned probability path to be learned via discrete flow matching, whose trajectories resemble acoustically plausible intermediate hypotheses~\cite{navon2025drax}. Finally, diffusion language models have been applied to ASR hypothesis rescoring and CTC hypothesis denoising~\cite{naveriani2026dlm}. However, because these models are trained to begin decoding from fully masked or completely random sequences, applying them to CTC outputs requires carefully tuning the optimal diffusion starting time to match the ``noise level'' of the initial CTC hypothesis.


\subsection{Edit-based sequence generation}

Edit-distance-based generation methods have been widely explored for text sequences~\cite{stern2019insertion,gu2019levenshtein,malmi2019lasertagger,mallinson2020felix}. Adapting these concepts to ASR, FastCorrect~\cite{leng2021fastcorrect} aligns an ASR hypothesis with its reference using edit distance, predicts the number of target tokens associated with each source token, and generates the corrected sequence in parallel. To improve robustness, it is pretrained on pseudo-pairs generated via deletion, insertion, and substitution noise, where the overall corruption rate and operation distribution are estimated directly from empirical ASR errors. Building on this, SoftCorrect~\cite{leng2023softcorrect} extends the framework by introducing soft error detection and a constrained CTC objective, which focuses the parallel correction process specifically on likely error positions. 
Finally, \cite{zhang2023patcorrect} augments the edit-based paradigm by incorporating explicit phonetic information from the initial hypothesis. Closely related to our work, NLE~\cite{dekel2026nle} refines an initial CTC hypothesis using an acoustically conditioned bidirectional LLM editor. It introduces interleaved insertion slots and trains the editor with a CTC-based latent-alignment objective.

\section{NAR refinement by Edit Flow}
\label{sec:method}

Given an input acoustic sequence $A$ and a token-level transcript $Y=(y_1,\ldots,y_N)$, we cast the recognition process as an iterative refinement of a first-pass CTC hypothesis. Figure~\ref{fig:method_overview} gives an overview of the proposed Edit Flow modeling procedure, which we detail below.


\begin{figure}[t]
\begin{tikzpicture}[
  scale=0.43, transform shape,
  font=\sffamily\large, 
  >=Stealth,
  bluebox/.style={rectangle, rounded corners, draw=blue!80!black, thick, fill=blue!5, align=center, minimum height=1.4cm, inner sep=3mm},
  tealbox/.style={rectangle, rounded corners, draw=teal, thick, fill=teal!5, align=center, minimum height=1.4cm, inner sep=3mm},
  graybox/.style={rectangle, rounded corners, draw=gray, thick, fill=gray!5, align=center, minimum height=1.4cm, inner sep=3mm},
  purplebox/.style={rectangle, rounded corners, draw=blue!60!magenta, thick, dashed, fill=blue!60!magenta!5, align=center, minimum height=1.7cm, inner sep=3mm},
  orangebox/.style={rectangle, rounded corners, draw=orange, thick, fill=orange!10, align=center, minimum height=1.0cm, inner sep=3mm}
]

\node[bluebox] (enc) at (0, 0) {\textbf{\textcolor{blue!80!black}{Acoustic Enc.}}\\ \textcolor{blue!80!black}{$H = \text{Enc}(A)$}};
\node[bluebox] (ctc) at (4.0, 0) {\textbf{\textcolor{blue!80!black}{CTC Head}}\\ \textcolor{blue!80!black}{$p_{ctc}(v|h_t)$}};
\node[tealbox] (hyp) at (8.0, 0) {\textbf{Collapsed}\\ \textbf{hypothesis} $x_0$};

\node[graybox] (ref) at (-3.2, -3) {\textbf{Reference}\\ $Y$};
\node[tealbox] (x1) at (0.8, -3) {$x_1 = [(\text{bos}); Y]$};
\node[tealbox] (align) at (8.0, -3) {\textbf{Levenshtein alignment}\\ $z_0, z_1$};
\node[tealbox] (sample) at (13.8, -3) {\textbf{sample} $z_t$\\ $x_t = \text{RemoveGap}(z_t)$};
\node[tealbox] (editflow) at (13.8, -5.5) {\textbf{Edit Flow}\\ $\lambda_{ins} \lambda_{sub} \lambda_{del} q_{ins} q_{sub}$};

\node[purplebox] (pretrain) at (0.0, -5.5) {
    \textcolor{blue!60!magenta}{\textbf{Edit-aware text pretraining}}\\
    \textcolor{black}{$z_0 \to z_1$ \textbf{by del/sub/ins}}\\
    \textcolor{black}{$H = 0$}
};

\node[orangebox, minimum width=11.5cm] (loss) at (9.35, -7.5) {$\mathcal{L} = \alpha\mathcal{L}_{CTC} + (1 - \alpha)\mathcal{L}_{EF}$};

\coordinate (audio_center) at (-3.8, 0);
\foreach \x/\h in {-0.4/0.3, -0.25/0.5, -0.1/0.8, 0.05/1.2, 0.2/0.7, 0.35/0.4, 0.5/0.2}
  \fill[gray!60!black] ($(audio_center)+(\x,-\h/2)$) rectangle ($(audio_center)+(\x+0.08,\h/2)$);
\draw[->, thick, densely dashed] ($(audio_center)+(0.6,0)$) -- (enc.west);

\draw[->, thick] (enc.east) -- (ctc.west);
\draw[->, thick] (ctc.east) -- (hyp.west);
\draw[->, thick] (ref.east) -- (x1.west);
\draw[->, thick] (x1.east) -- (align.west);

\draw[->, thick, teal] (hyp.south) -- (align.north);
\draw[->, thick, teal] (align.east) -- (sample.west);
\draw[->, thick, teal] (sample.south) -- (editflow.north);

\draw[->, thick] (enc.north) -- ++(0, 1.2) -| (sample.north);

\draw[->, thick, dashed, blue!60!magenta] (pretrain.east) -| (align.south);

\draw[->, thick, orange] (ctc.south) -- (ctc.south |- loss.north);
\draw[->, thick, orange] (editflow.south) -- (editflow.south |- loss.north);

\end{tikzpicture}
    \caption{
        Overview of the proposed CTC-seeded Edit Flow training procedure.
    }
    \label{fig:method_overview}
\end{figure}

\subsection{First-pass CTC}
\label{ssec:ctc_initialization}

CTC uses an acoustic encoder to map the input to a high-level representation
\[
    H=\operatorname{Enc}(A)=(h_1,\ldots,h_T).
\]
On top of $H$, a softmax layer produces frame-level posteriors
$p_{\mathrm{ctc}}(v\mid h_t)$ over the vocabulary $\mathcal{V}$ and the blank symbol
$\varnothing$. We obtain the greedy CTC alignment by selecting the most likely
symbol at each frame,
\[
    a_t
    =
    \arg\max_{v\in\mathcal{V}\cup\{\varnothing\}}
    p_{\mathrm{ctc}}(v\mid h_t),\quad t=1,\dots, T
\]
and then removing repetitions and blanks, denoted by the collapsing operator $\mathcal{B}$:
\[
    \hat{Y}^{(0)}=\mathcal{B}(a_{1:T}).
\]
The collapsed hypothesis supplies the initial token sequence, while the frame-level posteriors provide acoustic evidence for
guiding later edits.

\subsection{NAR refinement by Edit Flow}
\label{ssec:token_editflow}

\subsubsection{Variable-length edit path}
We adapt Edit Flows~\cite{havasi2026edit} to model the transformation from the
erroneous CTC hypothesis $\hat{Y}^{(0)}$ to the ground truth token sequence $Y$. Let
$\mathcal{V}_{+}=\mathcal{V}\cup\{\langle\mathrm{bos}\rangle\}$ and define
\[
    x_0=[\langle\mathrm{bos}\rangle;\hat{Y}^{(0)}],
    \qquad
    x_1=[\langle\mathrm{bos}\rangle;Y].
\]
Because $x_0$ and $x_1$ may have different lengths, we compute a Levenshtein alignment in an auxiliary space containing a gap symbol $\epsilon$. The aligned sequences satisfy
\[
    z_0,z_1\in(\mathcal{V}_{+}\cup\{\epsilon\})^M .
\]
Below is an example of this gap-augmented alignment.
\begin{table}[h]
\centering

\begin{tabular}{|c|c|c|c|}
\hline
\llap{$z_0$:\quad} WE
& SEES
& $\epsilon$
& KAT
\\
\hline
\llap{$z_1$:\quad} WE
& SEE
& A
& CAT
\\
\hline
\end{tabular}
\end{table}

Each aligned pair corresponds to an identity if the two symbols match, an insertion if $z_{0,m}=\epsilon$ and $z_{1,m}\neq\epsilon$, a deletion if $z_{0,m}\neq\epsilon$ and $z_{1,m}=\epsilon$, and a substitution if both symbols are non-empty but distinct.

For continuous refinement time $t\in[0,1]$, let $\kappa(t)$ be a monotonic noise scheduler with $\kappa(0)=0$, $\kappa(1)=1$, and $\dot{\kappa}(t)\geq0$ for $t \in (0,1)$. We sample an intermediate aligned state as
\[
    z_{t,m}
    =
    \begin{cases}
        z_{1,m}, & \text{with probability }\kappa(t),\\
        z_{0,m}, & \text{with probability }1-\kappa(t).
    \end{cases}
\]
The Edit Flow decoder receives as input the token sequence obtained after removing $\epsilon$ from $z_t$:
\[
    x_t=\operatorname{RemoveGap}(z_t).
\]
In other words, gaps help define the transformation path but never enter the decoder computation.

\subsubsection{Acoustic-conditioned edit field}
The Edit Flow decoder is a bidirectional transformer. Given the current token sequence $x_t$, flow time $t$, and acoustic memory $H$, it predicts quantities for the edit operations.

Let $\mathcal{P}_t$ denote all positions in $x_t$, including the protected
$\langle\mathrm{bos}\rangle$ position, and let
$\mathcal{I}_t=\{i\in\mathcal{P}_t:x_{t,i}\neq\langle\mathrm{bos}\rangle\}$
denote transcript-token positions. Insertion is defined for
$i\in\mathcal{P}_t$, while deletion and substitution are defined only for
$i\in\mathcal{I}_t$. The decoder predicts
\[
    \lambda_i^{\mathrm{ins}},\quad \text{for } i\in\mathcal{P}_t,
    \qquad
    \lambda_i^{\mathrm{sub}},\lambda_i^{\mathrm{del}},
    \quad \text{for } i\in\mathcal{I}_t,
\]
and token distributions
\[
    q_i^{\mathrm{ins}}(v),\quad v\in\mathcal{V},
    \qquad
    q_i^{\mathrm{sub}}(v),\quad v\in\mathcal{V}\setminus\{x_{t,i}\}.
\]
Deletion and substitution act on token $x_{t,i}$, while insertion acts on the
boundary after that token. The rate of a concrete edit proposal is
\[
u_\theta(e_i\mid x_t,H,x_0,t)=
\begin{cases}
\lambda_i^{\mathrm{ins}}q_i^{\mathrm{ins}}(v),
    & e_i=\operatorname{Ins}(v),\ i\in\mathcal{P}_t,\\
\lambda_i^{\mathrm{sub}}q_i^{\mathrm{sub}}(v),
    & e_i=\operatorname{Sub}(v),\ i\in\mathcal{I}_t,\\
\lambda_i^{\mathrm{del}},
    & e_i=\operatorname{Del},\ i\in\mathcal{I}_t.
\end{cases}
\]
Thus, the decoder edits directly in token space and changes sequence length without refining a frame-level alignment or expanding the sequence with editable padding positions.

\subsubsection{Edit Flow diffusion objective}
\label{sec:ef-diffusion-loss}
Let
\[
    \mathcal{M}_t
    =
    \{m\in\{1,\ldots,M\}:z_{t,m}\neq z_{1,m}\}
\]
index the remaining target edits in the aligned auxiliary space. Each
$m\in\mathcal{M}_t$ induces a target edit $e_m^\star$ at the corresponding visible token or boundary under $\operatorname{RemoveGap}$. Let
\[
    U_\theta(x_t,H,x_0,t)
    =
    \sum_{i\in\mathcal{P}_t}\lambda_i^{\mathrm{ins}}
    +
    \sum_{i\in\mathcal{I}_t}
    \left(
        \lambda_i^{\mathrm{sub}}
        +\lambda_i^{\mathrm{del}}
    \right)
\]
be the total edit intensity. The Edit Flow loss is
\begin{align*}
    \mathcal{L}_{\mathrm{EF}}
    =
    \mathbb{E}_{t,z_t}
    \bigg[ & 
        U_\theta(x_t,H,x_0,t)  \\
        & -
        \frac{\dot{\kappa}(t)}{1-\kappa(t)}
        \sum_{m\in\mathcal{M}_t}
        \log u_\theta(e_m^\star\mid x_t,H,x_0,t)
    \bigg].
\end{align*}

For supervised ASR, we linearly combine both CTC and Edit Flow losses to train all components end-to-end:
\[
    \mathcal{L}
    =
    \alpha\mathcal{L}_{\mathrm{CTC}}
    +(1-\alpha)\mathcal{L}_{\mathrm{EF}}
\]
where $\alpha$ is set to $0.3$ in this work.

\subsubsection{Edit-aware text pretraining}
\label{sec:pretraining}
Before supervised ASR training, we optionally pretrain the Edit Flow decoder on text-only data. Given a clean transcript $Y$, we construct a noisy source $\widetilde{Y}$ by independently applying deletion, substitution, and insertion corruptions to each token with probabilities $\rho_{\mathrm{del}}$, $\rho_{\mathrm{sub}}$, and $\rho_{\mathrm{ins}}$, respectively. 
The pair $(\widetilde{Y},Y)$ is aligned using the same Levenshtein auxiliary alignment and optimized with the standard Edit Flow objective. Unlike the acoustic-conditioned setting, the acoustic memory $H$ is set to $0$ during this pretraining stage. The purpose of this step is to provide a strong initialization for the large Edit Flow decoder, better equipping it to correct ASR-type errors. Empirically, we identify an optimal range for these noising probabilities that maximizes the model's performance.

\section{Inference Strategies}
\label{ssec:inference}



\subsection{Multi-step iterative refinement}
\label{sec:inference-refinement}

Operating as a diffusion model, Edit Flow utilizes $K$ steps to generate the final token sequence $x_1$ from the ($\langle\mathrm{bos}\rangle$-extended) initial CTC hypothesis $x_0$. At any given step $k$, with a step size $h_k$, we apply a parallel Poisson tau-leaping approximation to the continuous-time edit field~\cite{gillespie2001approximate}. This yields the insertion and delete-or-substitute event probabilities at every token position $i$:
\begin{align} \label{eq:inference-ins}
    p_{i,k}^{\mathrm{ins}}
    & =
    1-\exp(-h_k {\lambda}_{i,k}^{\mathrm{ins}}),
\\ \label{eq:inference-del-sub}
    p_{i,k}^{\mathrm{d/s}}
    & =
    1-\exp\!\left[
        -h_k\left(
        {\lambda}_{i,k}^{\mathrm{del}}
        +{\lambda}_{i,k}^{\mathrm{sub}}
        \right)
    \right].
\end{align}
For deterministic decoding, we accept an edit event if its probability exceeds an operation-specific threshold ($0.1$ in our experiments), subsequently applying the highest-scoring token edits at the accepted positions. Insertion and substitution tokens use the argmax of the token distributions ($q_i^{\text{ins}}(v)$ and $q_i^{\text{sub}}(v)$). In the stochastic decoding ablation, operation types and edit tokens are instead sampled according to their respective rates.
Empirically, deterministic decoding achieves superior performance, and we adopt it for all reported results.

In this work, we find that $K=2$ (with $h_k=0.5$) yields optimal performance in our experiments. That is, inference applies two parallel refinement rounds:
$
    \hat{Y}^{(0)}
    \longrightarrow
    \hat{Y}^{(1)}
    \longrightarrow
    \hat{Y}^{(2)}
$ where $\hat{Y}^{(2)}$ is returned as the final hypothesis.



\subsection{Audio classifier-free guidance (CFG)}

Classifier-free guidance (CFG,~\cite{ho2022classifier}) is a standard technique used to ensure that diffusion model generations strictly adhere to the conditioning signal.

We use CFG on the acoustic condition. During training, the acoustic memory $H$ is replaced by $0$ with a probability of 0.1, while the current sequence $x_t$ and initial sequence $x_0$ are kept unchanged. At inference, the audio-conditioned and
audio-dropped edit fields are combined with guidance scale $w$.
For operation $o\in\{\mathrm{ins},\mathrm{sub},\mathrm{del}\}$,
\[
    \widetilde{\lambda}_i^{o}
    =
    \exp\!\left(
        (1+w)
        \log\lambda_{i,\mathrm{audio}}^{o}
        -
        w
        \log\lambda_{i,\mathrm{drop}}^{o}
    \right).
\]
Insertion and substitution distributions are combined analogously: for $r\in\{\mathrm{ins},\mathrm{sub}\}$,
\[
    \widetilde{q}_i^{r}(v)
    \propto
    \exp\!\left(
        (1+w)
        \log q_{i,\mathrm{audio}}^{r}(v)
        -
        w
        \log q_{i,\mathrm{drop}}^{r}(v)
    \right),
\]
followed by normalization over the token vocabulary.
They are used in place of their counterparts in~\eqref{eq:inference-ins} and~\eqref{eq:inference-del-sub} for generation.

\subsection{CTC confidence guidance} 
When the initial CTC hypothesis is close to the reference transcript, unconstrained edit proposals risk rewriting tokens that already possess strong acoustic support. To mitigate this, we limit edits to acoustically uncertain regions. 

Because deletions and substitutions modify existing hypothesis tokens, these operations are gated by their corresponding token-level CTC confidence. Conversely, since insertions introduce new tokens between adjacent hypothesis tokens, they are gated by boundary-level CTC confidence. In both scenarios, low CTC confidence highlights locations where corrections are most likely to be beneficial.

During each refinement step, confidence scores are derived directly from the CTC forced alignment of the hypothesis (note that for the initial greedy CTC hypothesis, the greedy path coincides with the forced alignment). Let $\mathcal{R}_i$ denote the set of contiguous frames in the alignment corresponding to the repetitions of a non-blank token $\hat{y}_i$. We define the confidence of this token as
\[
    c_i
    =
    \frac{1}{|\mathcal{R}_i|}
    \sum_{t\in\mathcal{R}_i}
    p_{\mathrm{ctc}}(\hat{y}_i\mid h_t).
\]

We consider two definitions of insertion-boundary confidence.
The first, token-derived (TD) confidence, estimates the confidence of an insertion boundary from its adjacent token confidences. For a hypothesis of $N$ tokens, we define
\[
    b_j^{\mathrm{TD}}
    =
    \begin{cases}
        c_1, & j=0,\\
        \min(c_j,c_{j+1}),
            & 0<j<N,\\
        c_N, & j=N.
    \end{cases}
\]
The second, blank-derived (BD) confidence, directly uses the CTC blank posterior associated with the boundary. Let $B_j$ denote the set of frames assigned to the blank state at
boundary $j$ by the CTC alignment, plus one frame before and one frame after the blank frames. We define
\[
    b_j^{\mathrm{BD}}
    =
    \frac{1}{|B_j|}
    \sum_{t\in B_j}
    p_{\mathrm{ctc}}(\varnothing \mid h_t).
\] 
Both variants use the same token-level confidence $c_i$ to gate deletions and substitutions, and differ only in how confidence is assigned to insertion boundaries.
We consider different combinations of TD and BD for each refinement iteration.


Regardless of the iteration, these alignments supply the token confidence $c_i$ and boundary confidence $b_j$ necessary for the subsequent refinement. Guided by these metrics, an edit proposal is accepted only if it satisfies two criteria: its event probability (defined in~\eqref{eq:inference-ins} and~\eqref{eq:inference-del-sub}) must exceed the operation threshold ($0.1$ in our experiments), and the CTC confidence at the target position must fall below a designated threshold ($0.7$ in our experiments).


\section{Experiments}
\label{sec:experiments}


We evaluate our method on LibriSpeech~\cite{panayotov2015librispeech},
a standard benchmark of read English speech. We use the full 960-hour training set for ASR modeling, and the text-only (LM-training) data for pretraining the Edit Flow decoder. All models are evaluated on the standard LibriSpeech development and test sets, using word error rate (WER).


Our implementation is based on ESPnet~\cite{watanabe2018espnet} and uses a 5,000-unit SentencePiece vocabulary. Supervised ASR training is performed on a single NVIDIA H100 GPU with a batch size of 256 and the Noam learning-rate schedule. We train our model for 35 epochs and average models of the last 5 epochs as the final model for evaluation. For the acoustic encoder, we consider two configurations: an ESPNet transformer encoder trained on LibriSpeech from scratch, and a frozen pretrained Whisper encoder~\cite{radford2023robust}.

\subsection{Ablation Study}
\label{ssec:editflow_ablation}

We organize the ablation study around four factors that determine the behavior of our model.

\subsubsection{Model Capacity}

\begin{table}[t]
\centering
\caption{WERs (\%) obtained by different model sizes on the LibriSpeech dev sets.
Results are based on ESPNet encoder, without edit-aware pretraining. Two refinement steps are used for inference, without audio CFG or CTC confidence.}
\label{tab:size_ablation}
\vspace{-1ex}
\begin{tabular}{cccc}
\toprule
Model Size & Inference & dev-clean & dev-other \\
\hline\hline
Small & CTC & 4.0 & 9.3 \\
(125M) & Edit Flow & 3.8 & 9.0 \\ \hline
Medium & CTC & 3.8 & 8.8\\
(221M) & Edit Flow & 3.5 & 8.2 \\ \hline
Large & CTC & 3.6 & 8.3 \\
(403M) & Edit Flow & \textbf{3.4} & \textbf{7.9} \\
\bottomrule
\end{tabular}
\end{table}

In Table~\ref{tab:size_ablation}, we evaluate how scaling model capacity affects recognition performance. Small, medium, and large models are compared under identical settings: they are jointly trained with an ESPNet transformer encoder without 
pretraining, and inference utilizes two basic refinement steps without audio CFG or CTC confidence. 
The decoder sizes are 40M, 84M, and 155M parameters, paired with ESPNet encoders of 83M, 134M, and 244M parameters, respectively. Including the CTC head, the corresponding models contain 125M, 221M, and 403M parameters. 
We observe that increasing model capacity consistently improves development-set WER. The small model obtains a 3.8/9.0 WER on dev-clean/dev-other, which the medium model reduces to 3.5/8.2. 
The large model achieves the best result at 3.4/7.9. With the basic decoding strategy, edit flow consistently yields moderate WER improvements over jointly trained CTC.


\subsubsection{Edit-Aware Text Pretraining}

\begin{table}[t]
\centering
\caption{Effect of edit-aware pretraining strength. We use the large model and the same basic inference strategy as in Table~\ref{tab:size_ablation}.}
\label{tab:pretrain_base_ablation}
\vspace{-1ex}
\begin{tabular}{lcc}
\toprule
Edit-noise $\rho$ & dev-clean & dev-other \\
\midrule
No Pretrain & 3.4 & 7.9 \\
0.01 & 3.5 & 8.1 \\
0.03 & \textbf{3.3} & \textbf{7.5} \\
0.05 & 3.3 & 7.6 \\
0.07 & 3.4 & 7.7 \\
\bottomrule
\end{tabular}
\end{table}

\begin{table}[t]
\centering
\caption{WERs by large edit-aware pretrained decoder using CTC confidence (TD and BD denote token-derived and blank-derived insertion-boundary confidence, respectively.). No audio CFG is used.}
\label{tab:inference_ablation}
\vspace{-1ex}
\begin{tabular}{llcc}
\toprule
Setting & Value & dev-clean & dev-other \\
\midrule
\multicolumn{4}{l}{\emph{Number of refinement steps, with TD$\rightarrow$BD confidence}} \\
\quad Steps & 0 (CTC) &   3.6 & 8.3 \\
\quad Steps & 1 & 3.1 & 7.3 \\
\quad Steps & 2 & \textbf{2.7} & \textbf{6.8} \\
\quad Steps & 4 & 2.8 & 7.0 \\
\midrule
\multicolumn{4}{l}{\emph{CTC confidence guidance with two refinement steps}} \\
\quad Guide & Off
    & 3.3 & 7.5 \\
\quad Guide & $\mathrm{TD}\!\rightarrow\!\mathrm{TD}$
    & 2.9 & 7.1 \\
\quad Guide & $\mathrm{BD}\!\rightarrow\!\mathrm{BD}$
    & 3.0 & 7.3 \\
\quad Guide & $\mathrm{BD}\!\rightarrow\!\mathrm{TD}$
    & 3.2 & 7.2 \\
\quad Guide & $\mathrm{TD}\!\rightarrow\!\mathrm{BD}$
    & \textbf{2.7} & \textbf{6.8} \\
\bottomrule
\end{tabular}
\end{table}

\begin{table}[t]
\centering
\caption{Effect of audio CFG scale on dev WERs. We use pretrained large decoder with the optimal CTC confidence setup.}
\label{tab:audio_cfg}
\vspace{-1ex}
\begin{tabular}{ccccc}
\toprule
Scale &
\multicolumn{2}{c}{ESPNet Encoder} &
\multicolumn{2}{c}{Whisper Base Enc. (25M)} \\
\cmidrule(lr){2-3}
\cmidrule(lr){4-5}
$w$ & dev-clean & dev-other & dev-clean & dev-other \\
\midrule
0.00 & 2.7 & 6.8 & 2.3 & 5.6 \\
0.10 & 2.5 & 6.1 & 2.3 & 5.2 \\
0.30 & \textbf{2.3} & \textbf{5.5} & \textbf{2.1} & \textbf{4.8} \\
0.50 & 2.5 & 5.8 & \textbf{2.1} & 4.9 \\
0.70 & 2.6 & 6.4 & 2.2 & 5.1 \\
\bottomrule
\end{tabular}
\end{table}

From now on, we focus our evaluation on the large decoder. 
Table~\ref{tab:pretrain_base_ablation} investigates the impact of the corruption strength $\rho=\rho_{\mathrm{del}}=\rho_{\mathrm{sub}}=\rho_{\mathrm{ins}}$ during text-only Edit Flow pretraining (see Sec.~\ref{sec:pretraining} for details). For this ablation, we employ the large model and follow the same inference strategies as in Table~\ref{tab:size_ablation}. Without decoder pretraining, the model achieves a WER of 3.4/7.9. A noise strength of $\rho=0.03$ yields the best performance, improving the WERs to 3.3/7.5 with sizable improvement on dev-other. We hypothesize that the optimal corruption level depends heavily on the accuracy of the baseline CTC hypotheses, as matching this noise level tightly aligns the pretraining distribution with the actual fine-tuning conditions. Therefore, we adopt $\rho=0.03$ for edit-aware pretraining below.

\begin{figure*}[!t]
\centering
\begin{tabular}{|p{0.1\textwidth}@{\hspace{0.00\linewidth}}|p{0.28\textwidth}|p{0.30\textwidth}|p{0.23\textwidth}|}
\hline
& Example 1 & Example 2 & Example 3 \tabularnewline
\hline
\raggedright CTC &
\raggedright no it is that from curiositythes from love &
\raggedright i ohll hang it uncle why can't you let me alone &
\raggedright but his mother huggged them close \tabularnewline
\hline
\raggedright Step 1 &
\raggedright no it is \textcolor{red}{\bf not} from \textcolor{red}{\bf curiosity} \textcolor{red}{\bf it} from love &
\raggedright ohll hang it uncle why can't you let me alone &
\raggedright but his mother \textcolor{red}{\bf hugged} \textcolor{red}{\bf hem} close \tabularnewline
\hline
\raggedright Step 2 &
\raggedright no it is not from curiosity it \textcolor{red}{\bf is} from love &
\raggedright \textcolor{red}{\bf oh} hang it uncle why can't you let me alone &
\raggedright but his mother hugged \textcolor{red}{\bf him} close \tabularnewline
\hline
\end{tabular}
\caption{Examples of two-step Edit Flow refinement. For all examples, step 2 decoding results match ground truth transcripts.}
\label{fig:edit_examples}
\end{figure*}

\subsubsection{CTC confidence}
Table~\ref{tab:inference_ablation} investigates the inference-time behavior of the large decoder equipped with edit-aware pretraining. The first block evaluates the impact of varying the number of refinement steps. We use TD confidence for the first refinement step and BD confidence for subsequent steps. We observe that performance peaks at two refinement steps.
The second block compares different insertion-boundary confidence
strategies under this optimal two-step setting. TD derives the
boundary confidence from adjacent token confidences, whereas BD
uses the CTC blank posterior associated with the boundary.
Among the evaluated schedules, TD$\rightarrow$BD achieves the best WERs, and we therefore adopt it as our default inference configuration. Notably, the combination of edit-aware pretraining and CTC confidence guidance improves the large model's WERs of 3.4/7.9 (see Table~\ref{tab:size_ablation}) to 2.7/6.8.

\subsubsection{Audio Classifier-Free Guidance}

Table~\ref{tab:audio_cfg} evaluates the impact of audio classifier-free guidance (CFG) using both the ESPNet and Whisper Base encoders. For these experiments, we pair the pretrained large decoder with the optimal CTC confidence setup established above. During inference, the conditioned and unconditioned predictions are combined using an audio-CFG scale $w$, where $w=0$ corresponds to disabled guidance. For the ESPNet encoder, increasing $w$ from 0.00 to 0.30 reduces the WER from 2.7/6.8 to 2.3/5.5 on dev-clean/dev-other. However, applying larger scales diminishes these gains. A similar trend is observed with the Whisper Base encoder, where $w=0.30$ yields the best overall WER of 2.1/4.8. Thus we adopt $w=0.30$ for 
final evaluations.

\subsection{Final Test Results}
\label{ssec:final_results}


Table~\ref{tab:final_test_results} presents  test WERs using inference hyperparameters optimized on development sets. For each acoustic encoder, we report Edit Flow results using the large decoder, evaluating performance both with and without edit-aware pretraining.

For the ESPNet encoder, the CTC baseline yields 3.5/8.1 WER on test-clean/test-other. Edit Flow refinement without pretraining drops this to 2.9/6.6, and edit-aware pretraining further reduces it to 2.6/5.8. This confirms that the CTC-seeded variable-length editing actively corrects first-pass residual errors, rather than relying on a stronger encoder. The frozen Whisper Base encoder follows an identical trend: the 3.1/6.9 CTC baseline improves to 2.4/5.7 without pretraining, and reaches 2.2/5.1 with the pretrained decoder.
Using a frozen Whisper Medium encoder (308M) further drives test WERs down to 2.0/4.7. 
Across all configurations, our method significantly outperforms the CTC baselines, with decoder pretraining yielding further relative WER reductions compared to non-pretrained models.
In Figure~\ref{fig:edit_examples} we provide examples of our decoding process.
Furthermore, we evaluate the inference RTF on LibriSpeech test-other using a single NVIDIA H100 with batch size one. 
Our method achieves a Real-Time Factor (RTF) of ~0.0044, an order of magnitude faster than prior approaches (baseline RTFs are cited directly from their respective papers and were evaluated on different hardware).

We also compare our approach against prior non-autoregressive and diffusion-based ASR systems under comparable settings. Our best system achieves competitive performance while requiring only a moderate model size and two parallel refinement steps. Crucially, it accomplishes this without relying on strong supervised initialization or massive external datasets (e.g., MDM-ASR~\cite{yen2026mdmasr} is initialized from a large supervised AED model, and Drax~\cite{navon2025drax} leverages 15K hours of multilingual training data).

\begin{table}[t]
\centering
\caption{
WERs (\%) on the LibriSpeech test sets.
RTF is measured on test-other with batch size 1.
}
\label{tab:final_test_results}
\vspace{-1ex}
\begin{tabular}{@{}l@{\hspace{0.01\linewidth}}l@{\hspace{0.02\linewidth}}l@{\hspace{0.01\linewidth}}r@{\hspace{0.02\linewidth}}r@{\hspace{0.02\linewidth}}c@{}}
\toprule
Method
& Inference
& Pretrain
& test-clean
& test-other
& RTF $\downarrow$ \\
\midrule

\multicolumn{3}{l}{ESPNet large transducer, no fusion}
    & 2.9 & 7.1 & -- \\
\multicolumn{3}{l}{TransFusion~\cite{baas2022transfusion}}
    & 6.7 & 8.8 & -- \\
\multicolumn{3}{l}{CASS-NAT~\cite{fan2023cassnat}}
    & 3.8 & 9.1 & 0.010 \\
\multicolumn{3}{l}{FDDM~\cite{yeh2024crossmodal}}
    & 4.0 & 7.2 & -- \\
\multicolumn{3}{l}{Whisper-LLaDA~\cite{wang2025audio} (non deliberation)}
    & 2.8 & 5.8 & 0.194 \\
\multicolumn{3}{l}{Drax~\cite{navon2025drax} (NFE=16, w.o. ensemble)}
    & 2.6 & 5.7 & 0.0311 \\
\multicolumn{3}{l}{dLLM-ASR~\cite{tian2026dllmasr}}
    & 2.3 & 5.2 & 0.076 \\
\multicolumn{3}{l}{MDM-ASR~\cite{yen2026mdmasr}}
    & 1.8 & 3.6 & 0.0214 \\

\midrule
\multicolumn{6}{l}{\textit{Ours}} \\
\midrule

\multirow{4}{*}{
    \parbox{8em}{\centering
    ESPNet Encoder\\
    (model size: 403M)}
}
& CTC
& \multirow{2}{*}{No}
& 3.5 & 8.1 & 0.0019 \\
& Edit Flow
&
& 2.9 & 6.6 & 0.0081 \\

\cmidrule(lr){2-5}

& CTC
& \multirow{2}{*}{Yes}
& 3.5 & 7.9 & \\
& Edit Flow
&
& \textbf{2.6} & \textbf{5.8} & \\

\midrule

\multirow{4}{*}{
    \parbox{8em}{\centering
    Whisper Base\\
    (model size: 179M)}
}
& CTC
& \multirow{2}{*}{No}
& 3.1 & 6.9 & 0.0008 \\
& Edit Flow
&
& 2.4 & 5.7 & 0.0044 \\

\cmidrule(lr){2-5}

& CTC
& \multirow{2}{*}{Yes}
& 2.9 & 6.7 & \\
& Edit Flow
&
& \textbf{2.2} & \textbf{5.1} & \\

\midrule

\multirow{4}{*}{
    \parbox{8em}{\centering
    Whisper Medium\\
    (model size: 467M)}
}
& CTC
& \multirow{2}{*}{No}
& 2.7 & 6.5 & 0.0010 \\
& Edit Flow
&
& 2.1 & 5.4 & 0.0044 \\

\cmidrule(lr){2-5}

& CTC
& \multirow{2}{*}{Yes}
& 2.6 & 6.1 & \\
& Edit Flow
&
& \textbf{2.0} & \textbf{4.7} & \\

\bottomrule
\end{tabular}
\end{table}
\section{Conclusion}
\label{sec:conclusion}


In this work, we have introduced a non-autoregressive ASR model that frames decoding as a token-level edit refinement of an initial CTC hypothesis, departing from traditional generation-from-noise approaches. Our decoder executes parallel insertions, deletions, and substitutions through an acoustic-conditioned Edit Flow, guided by a two-stage confidence mechanism. Additionally, we designed an edit-aware pretraining scheme to learn correction behaviors from text-only data, validating the entire framework on both ESPnet and frozen Whisper encoders. Ultimately, our findings demonstrate that CTC-seeded, variable-length editing is a powerful and efficient alternative to fully masked generation, requiring minimal parallel correction steps. Future work will explore  expanding this framework to multilingual speech recognition.

\section*{Acknowledgments}
Parts of this manuscript were drafted with assistance from AI. The entire codebase and the paper were reviewed and edited by the authors.

\newpage

\bibliographystyle{IEEEtran}
\bibliography{mybib}

@inproceedings{higuchi2020maskctc,
  title     = {{Mask CTC: Non-Autoregressive End-to-End ASR with CTC and Mask Predict}},
  author    = {Higuchi, Yosuke and Watanabe, Shinji and Chen, Nanxin and Ogawa, Tetsuji and Kobayashi, Tetsunori},
  booktitle = {{Proceedings of Interspeech 2020}},
  pages     = {3655--3659},
  year      = {2020},
  doi       = {10.21437/Interspeech.2020-2404}
}

@inproceedings{chan2020imputer,
  author    = {Chan, William and Saharia, Chitwan and Hinton, Geoffrey and Norouzi, Mohammad and Jaitly, Navdeep},
  title     = {Imputer: Sequence Modelling via Imputation and Dynamic Programming},
  booktitle = {Proceedings of the 37th International Conference on Machine Learning},
  pages     = {1403--1413},
  year      = {2020},
  volume    = {119},
  publisher = {PMLR},
}

@inproceedings{ghazvininejad2019maskpredict,
  author    = {Ghazvininejad, Marjan and Levy, Omer and Liu, Yinhan and Zettlemoyer, Luke},
  title     = {Mask-Predict: Parallel Decoding of Conditional Masked Language Models},
  booktitle = {EMNLP-IJCNLP},
  year      = {2019},
}

@article{wang2021deliberation,
  author  = {Wang, Weiran and Hu, Ke and Sainath, Tara N.},
  title   = {Deliberation of Streaming RNN-Transducer by Non-autoregressive Decoding},
  journal = {arXiv preprint arXiv:2112.11442},
  year    = {2021},
}

@inproceedings{watanabe2018espnet,
  author    = {Watanabe, Shinji and Hori, Takaaki and Karita, Shigeki and Hayashi, Tomoki and Nishitoba, Jiro and Unno, Yuya and Soplin, Nelson Enrique Yalta and Heymann, Jahn and Wiesner, Matthew and Chen, Nanxin and Renduchintala, Adithya and Ochiai, Tsubasa},
  title     = {{ESPnet}: End-to-End Speech Processing Toolkit},
  booktitle = {Interspeech},
  year      = {2018},
}

@inproceedings{panayotov2015librispeech,
  author    = {Panayotov, Vassil and Chen, Guoguo and Povey, Daniel and Khudanpur, Sanjeev},
  title     = {{LibriSpeech}: An {ASR} Corpus Based on Public Domain Audio Books},
  booktitle = {ICASSP},
  year      = {2015},
}

@inproceedings{wang2022streamingalignrefine,
  author    = {Wang, Weiran and Hu, Ke and Sainath, Tara N.},
  title     = {Streaming Align-Refine for Non-autoregressive Deliberation},
  booktitle = {Interspeech},
  year      = {2022},
}

@inproceedings{graves2006ctc,
  title={Connectionist temporal classification: labelling unsegmented sequence data with recurrent neural networks},
  author={Graves, Alex and Fern{\'a}ndez, Santiago and Gomez, Faustino and Schmidhuber, J{\"u}rgen},
  booktitle={Proceedings of the 23rd international conference on Machine learning},
  year={2006}
}

@misc{graves2012rnnt,
  author = {Alex Graves},
  title = {Sequence Transduction with Recurrent Neural Networks},
  year = {2012},
  eprint = {1211.3711},
  archivePrefix = {arXiv},
}

@inproceedings{chan2016las,
  author = {William Chan and Navdeep Jaitly and Quoc V. Le and Oriol Vinyals},
  title = {Listen, Attend and Spell: A Neural Network for Large Vocabulary Conversational Speech Recognition},
  booktitle = {2016 IEEE International Conference on Acoustics, Speech and Signal Processing},
  pages = {4960--4964},
  year = {2016},
}

@inproceedings{higuchi2021improvedmaskctc,
  author = {Yosuke Higuchi and Hirofumi Inaguma and Shinji Watanabe and Tetsuji Ogawa and Tetsunori Kobayashi},
  title = {Improved Mask-{CTC} for Non-Autoregressive End-to-End {ASR}},
  booktitle = {2021 IEEE International Conference on Acoustics, Speech and Signal Processing},
  year = {2021},
}

@inproceedings{fujita2020insertionasr,
  author = {Yuya Fujita and Shinji Watanabe and Motoi Omachi and Xuankai Chang},
  title = {Insertion-Based Modeling for End-to-End Automatic Speech Recognition},
  booktitle = {Interspeech 2020},
  pages = {3660--3664},
  year = {2020},
}

@ARTICLE{fan2023cassnat,
  author={Fan, Ruchao and Chu, Wei and Chang, Peng and Alwan, Abeer},
  journal={IEEE/ACM Transactions on Audio, Speech, and Language Processing}, 
  title={A CTC Alignment-Based Non-Autoregressive Transformer for End-to-End Automatic Speech Recognition}, 
  year={2023},
  }

@inproceedings{chi2021align,
  title={Align-refine: Non-autoregressive speech recognition via iterative realignment},
  author={Chi, Ethan A and Salazar, Julian and Kirchhoff, Katrin},
  booktitle={Proceedings of the 2021 Conference of the North American Chapter of the Association for Computational Linguistics: Human Language Technologies},
  pages={1920--1927},
  year={2021}
}

@misc{baas2022transfusion,
  author = {Matthew Baas and Kevin Eloff and Herman Kamper},
  title = {TransFusion: Transcribing Speech with Multinomial Diffusion},
  year = {2022},
  eprint = {2210.07677},
  archivePrefix = {arXiv},
}

@misc{yen2026mdmasr,
  author = {Hao Yen and Pin-Jui Ku and Ante Jukic and Sabato Marco Siniscalchi},
  title = {{MDM-ASR}: Bridging Accuracy and Efficiency in {ASR} with Diffusion-Based Non-Autoregressive Decoding},
  year = {2026},
  eprint = {2602.18952},
  archivePrefix = {arXiv},
}

@misc{tian2026dllmasr,
  author = {Wenjie Tian and Bingshen Mu and Guobin Ma and Xuelong Geng and Zhixian Zhao and Lei Xie},
  title = {{dLLM-ASR}: A Faster Diffusion {LLM}-Based Framework for Speech Recognition},
  year = {2026},
  eprint = {2601.17902},
  archivePrefix = {arXiv},
}

@misc{navon2025drax,
  author = {Aviv Navon and Aviv Shamsian and Neta Glazer and Yael Segal-Feldman and Gill Hetz and Joseph Keshet and Ethan Fetaya},
  title = {Drax: Speech Recognition with Discrete Flow Matching},
  year = {2025},
}

@inproceedings{leng2021fastcorrect,
  author = {Yichong Leng and Xu Tan and Linchen Zhu and Jin Xu and Renqian Luo and Linquan Liu and Tao Qin and Xiang-Yang Li and Ed Lin and Tie-Yan Liu},
  title = {FastCorrect: Fast Error Correction with Edit Alignment for Automatic Speech Recognition},
  booktitle = {Advances in Neural Information Processing Systems},
  volume = {34},
  pages = {21708--21719},
  year = {2021},
}

@article{leng2023softcorrect,
  author = {Yichong Leng and Xu Tan and Wenjie Liu and Kaitao Song and Rui Wang and Xiang-Yang Li and Tao Qin and Ed Lin and Tie-Yan Liu},
  title = {SoftCorrect: Error Correction with Soft Detection for Automatic Speech Recognition},
  journal = {Proceedings of the AAAI Conference on Artificial Intelligence},
  volume = {37},
  number = {11},
  pages = {13034--13042},
  year = {2023},
}

@inproceedings{futami2022pcmlm,
  author = {Hayato Futami and Hirofumi Inaguma and Sei Ueno and Masato Mimura and Shinsuke Sakai and Tatsuya Kawahara},
  title = {Non-Autoregressive Error Correction for {CTC}-Based {ASR} with Phone-Conditioned Masked {LM}},
  booktitle = {Interspeech 2022},
  pages = {3889--3893},
  year = {2022},
}

@inproceedings{zhang2023patcorrect,
  author = {Ziji Zhang and Zhehui Wang and Rajesh Kamma and Sharanya Eswaran and Narayanan Sadagopan},
  title = {{PATCorrect}: Non-Autoregressive Phoneme-Augmented Transformer for {ASR} Error Correction},
  booktitle = {Interspeech 2023},
  pages = {3904--3908},
  year = {2023},
}

@misc{dekel2026nle,
  author = {Avihu Dekel and Samuel Thomas and Takashi Fukada and George Saon},
  title = {{NLE}: Non-Autoregressive {LLM}-Based {ASR} by Transcript Editing},
  year = {2026},
  eprint = {2603.08397},
  archivePrefix = {arXiv},
}

@article{havasi2026edit,
  title={Edit Flows: Flow Matching with Edit Operations},
  author={Havasi, Marton and Karrer, Brian and Gat, Itai and Chen, Ricky TQ},
  journal={arXiv preprint arXiv:2506.09018},
  year={2025}
}

@inproceedings{radford2023robust,
  author = {Alec Radford and Jong Wook Kim and Tao Xu and Greg Brockman and Christine McLeavey and Ilya Sutskever},
  title = {Robust Speech Recognition via Large-Scale Weak Supervision},
  booktitle = {Proceedings of the 40th International Conference on Machine Learning},
  series = {Proceedings of Machine Learning Research},
  volume = {202},
  pages = {28492--28518},
  year = {2023},
}

@article{wang2025audio,
  title={Audio-Conditioned Diffusion LLMs for ASR and Deliberation Processing},
  author={Wang, Mengqi and Liu, Zhan and Jin, Zengrui and Sun, Guangzhi and Zhang, Chao and Woodland, Philip C.},
  journal={arXiv preprint arXiv:2509.16622},
  year={2025}
}

@article{naveriani2026dlm,
  title={Diffusion Language Models for Speech Recognition},
  author={Naveriani, Davyd and Zeyer, Albert and Schluter, Ralf and Ney, Hermann},
  journal={arXiv preprint arXiv:2604.14001},
  year={2026}
}

@inproceedings{stern2019insertion,
  title = {Insertion Transformer: Flexible Sequence Generation via Insertion Operations},
  author = {Stern, Mitchell and Chan, William and Kiros, Jamie and Uszkoreit, Jakob},
  booktitle = {Proceedings of the 36th International Conference on Machine Learning},
  pages = {5976--5985},
  year = {2019},
  volume = {97},
  series = {Proceedings of Machine Learning Research},
  publisher = {PMLR},
}

@inproceedings{gu2019levenshtein,
  title = {Levenshtein Transformer},
  author = {Gu, Jiatao and Wang, Changhan and Zhao, Junbo},
  booktitle = {Advances in Neural Information Processing Systems},
  volume = {32},
  year = {2019},
}

@inproceedings{malmi2019lasertagger,
  title = {Encode, Tag, Realize: High-Precision Text Editing},
  author = {Malmi, Eric and Krause, Sebastian and Rothe, Sascha and Mirylenka, Daniil and Severyn, Aliaksei},
  booktitle = {Proceedings of EMNLP-IJCNLP},
  pages = {5054--5065},
  year = {2019},
}

@inproceedings{mallinson2020felix,
  title = {{FELIX}: Flexible Text Editing Through Tagging and Insertion},
  author = {Mallinson, Jonathan and Severyn, Aliaksei and Malmi, Eric and Garrido, Guillermo},
  booktitle = {Findings of the Association for Computational Linguistics: EMNLP 2020},
  pages = {1244--1255},
  year = {2020},
}

@article{gillespie2001approximate,
  title={Approximate accelerated stochastic simulation of chemically reacting systems},
  author={Gillespie, Daniel T.},
  journal={The Journal of Chemical Physics},
  volume={115},
  number={4},
  pages={1716--1733},
  year={2001},
}

@inproceedings{yeh2024crossmodal,
  title={Cross-Modality Diffusion Modeling and Sampling for Speech Recognition.},
  author={Yeh, Chia-Kai and Chen, Chih-Chun and Hsu, Ching-Hsien and Chien, Jen-Tzung},
  booktitle={INTERSPEECH},
  year={2024}
}

@inproceedings{nie2025llada,
 author = {Nie, Shen and Zhu, Fengqi and You, Zebin and Zhang, Xiaolu and Ou, Jingyang and Hu, Jun and Zhou, Jun and Lin, Yankai and Wen, Ji-Rong and LI, Chongxuan},
 booktitle = {Advances in Neural Information Processing Systems},
 pages = {50608--50646},
 publisher = {Curran Associates, Inc.},
 title = {Large Language Diffusion Models},
 year = {2025}
}

@article{puvvada2024less,
  title={Less is more: Accurate speech recognition \& translation without web-scale data},
  author={Puvvada, Krishna C and {\.Z}elasko, Piotr and Huang, He and Hrinchuk, Oleksii and Koluguri, Nithin Rao and Dhawan, Kunal and Majumdar, Somshubra and Rastorgueva, Elena and Chen, Zhehuai and Lavrukhin, Vitaly and others},
  journal={arXiv preprint arXiv:2406.19674},
  year={2024}
}

@article{austin2021structured,
  title={Structured denoising diffusion models in discrete state-spaces},
  author={Austin, Jacob and Johnson, Daniel D and Ho, Jonathan and Tarlow, Daniel and Van Den Berg, Rianne},
  journal={Advances in neural information processing systems},
  volume={34},
  pages={17981--17993},
  year={2021}
}

@article{li2022diffusion,
  title={Diffusion-lm improves controllable text generation},
  author={Li, Xiang and Thickstun, John and Gulrajani, Ishaan and Liang, Percy S and Hashimoto, Tatsunori B},
  journal={Advances in neural information processing systems},
  volume={35},
  pages={4328--4343},
  year={2022}
}

@article{ho2022classifier,
  title={Classifier-free diffusion guidance},
  author={Ho, Jonathan and Salimans, Tim},
  journal={arXiv preprint arXiv:2207.12598},
  year={2022}
}

\end{document}